\documentclass[a4paper, astronm]{article}

\RequirePackage{pifont,latexsym,ifthen,theorem,rotating,calc}
\RequirePackage{amsfonts,amssymb,amsbsy,amsmath}

\usepackage{graphicx}
\usepackage{astronm}

\newcommand{\ud}{\mathrm{d}}

\title{ Dynamic Modeling in Health Research as a framework
for developing statistical applications free of misuse of statistics}

\author{Vladislav Moltchanov \\
\\
National Institute for Health and Welfare (THL)\\
Department of Chronic Disease Prevention \\
P.O. Box 30
FI-00271 Helsinki
Finland \\
\emph{email}: vladislav.moltchanov@thl.fi }

\begin{document}

\maketitle

\begin{abstract} We introduce a novel framework for developing statistical applications in health research, based on dynamic modeling of the investigated processes. We formulate the principles of dynamic modeling in health research, which  are coherent to those in other fields of research. Dynamic models explicitly describe causal relations which are to be adequately accounted in statistical methods, making them free of misuse of statistics and statistical fallacy.

We propose the Dynamic Model of Population Health describing temporal changes in health indicators, having nature of state variables. The Dynamic Regression Method  was developed as statistical method for  the identification of the model. This method evaluates cohort trends for  state variables at each age and calendar year. The method is illustrated by evaluating cohort trends for the Body Mass Index for men, using survey data collected in the years 1982, 1987, 1992, in North Karelia, Finland.

\end{abstract}



%


\section{Introduction. \label{section intro}}

Misuse of statistics and statistical fallacy are issues of concern in many fields, including medical research. The detailed  classification of misuse and recommendations of how to avoid statistical fallacy could be found in books \cite{Jaffe}, \cite{Campbell}.
One category of misuse, "lack of knowledge on subject matter" \cite{Jaffe}, could  well be interpreted as addressing  causality among other things.

Recently, it was acknowledged that  a large proportion of published
medical research contains statistical errors and
shortcomings . "The problem is a serious
one, as the inappropriate use of statistical analysis
may lead to incorrect conclusions, artificial research
results and a waste of valuable resources" \cite{Strasak}. Interestingly, the  authors believe  that one of the reason for misuse is lack of statistical literacy:
"Medical
researchers have to be encouraged to learn more
about statistics, as various studies point to a lack of
statistical knowledge among medical residents"
and
"statisticians should be involved
early in study design"

In disagreement with this, we consider the case  of statistical misuse, which occurs when a formally correct statistical method is used, however, causality is missing. These are the methods of  evaluating secular trends in health indicators using data from a set of independent cross-sectional surveys (review on these methods is presented in section \ref{histtta}). The linear secular trends are  also used  as a tool for interecensal estimates of population size (review of this could be found in  \cite{Moltchanov1999}. We use term "secular trends" to refer to all these methods.

Formally, methods evaluating secular trends look correct: data, linear model and assumptions on data properties, in combination lead to evaluation of the model's parameters and testing hypotheses.  In this schema, the wrong element is assumptions on data, usually suggesting   smooth-line type of dependency of estimated means.

Note that Hill's criteria for causation (see, for example,
{http://en.wikipedia.org/wiki/Bradford-Hill\_criteria}  ), though sound reasonable, provide only circumstantial evidence for causality,  and leave plenty of room for subjective judgments.

\begin{figure}
\begin{center}
\centerline{\includegraphics[angle=90,width=20.5cm]{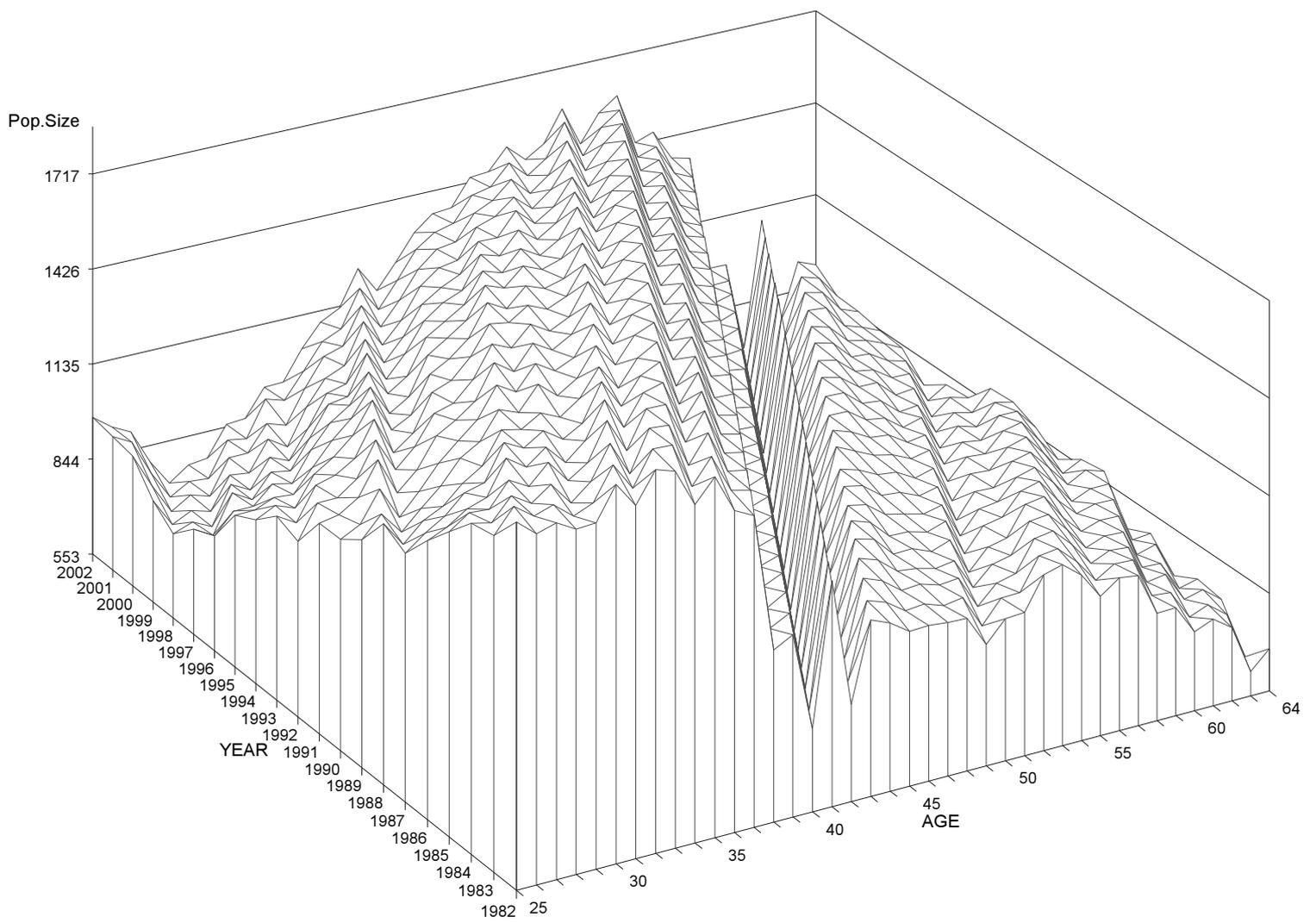}}
\end{center}
\caption{Population size, Men, Original data. Count by year and age.
\label{pops}}
\end{figure}

To our view, the data on population size shown on Figure \ref{pops} provide a very strong evidence that secular trends do not exist in nature. Rather  there are smooth evolution of population size along birth cohort. However, some of the statisticians and field researcher may disagree with this statement.

To  address properly and unambiguously causality in a real world process, first of all, knowledge and skill on dynamic  modeling  should be applied. Only at the next stage, statistical tools are to be considered to evaluate parameters of the dynamic model.

 Our practical task (target task) is to evaluate temporal changes in health on population level, using data from a set of independent cross-sectional population surveys. Traditionally, this task is approached by evaluating secular trends, which appeared to be a statistical fallacy. To build up an alternative approach, we develop Dynamic Model of Population Health (DMPH) and
statistical method for  its identification, the Dynamic Regression Method(DRM),  producing time trends for health-related indicators within birth cohorts (Cohort trends, or C-trends for short).

In turn, to build up a dynamic model, first we derive some principles, which we call the Principles of Dynamic Modeling in Health Research. These Principles are independent of the target task, so they could be applied to any other task, for example, to follow-up  analysis with end-points.

The aims of this paper are as follows:
\begin{itemize}
\item  to derive the Principles of Dynamic Modeling In Health Research
\item to develop the Dynamic Model of Population Health
\item to build up the Dynamic Regression Method and algorithm
\item to run the illustrative analysis
\end{itemize}

Note that each of the four parts above is worth of more detailed, separate  presentation.

Therefore, the challenge was to provide concise and logically completed descriptions of all parts, clearly outlining logical interrelations between them.

The earlier version of the Dynamic Regression Method was developed and presented  by \cite{MolMic}, where C - trends were suggested as an alternative to circular trends, used so far.
Historically, the  method developed in MONICA for checking
consistency of the reported demographic data (\cite{Moltchanov1999})
served  as the prototype for the method developed by \cite{MolMic}.
Some general aspects, such as criteria  for commonly used health indicators to serve as  system State Variables, have been considered earlier by
\cite{Moltchanov1993}.

Section \ref{histtta}  contains historical review of time-trend analysis.
Section \ref{prin}  presents history and key properties of Dynamic Modeling.
Section \ref{ModPopLev} describes  Dynamic Model paradigm being applied to health research on Individual and Population Level.
In Section \ref{Anaform} we
present the analytical model for the  case of continuous, normally
distributed one parameter.
The example of application, employing the
data of real study, is given in section \ref{Exa}.
Section \ref{conclu} contains  conclusion and discussion.


\section{History of Time-Trend Analysis in Health Research\label{histtta}}
Time-trend analysis  of health indicators   has been  widely used so far in
health research. One motivation for this comes from  the fact that
running a cross-sectional survey is the cost-effective way to collect the
data for such an analysis \cite{Mann2003}.

So far, this  problem has been approached by assessing  trends over
time for means  and other statistics (for example, percentiles) of
the age-specific distributions of parameters of interest, such as
traditional risk factor indicators (for example, systolic and
diastolic blood pressure, cholesterol, body mass index), and their
categorical derivatives (such as prevalence of high blood pressure,
prevalence of high cholesterol, prevalence of obesity).

Various terms are used in  literature for such trends: "trends",
"secular trends", "time trends". Here we will  use the term "secular
trends" for all of them.

Among  approaches used for trend analysis, the first one, "trends by
linear regression", was the key element of the analysis in the WHO
MONICA Project \cite{MoniMono}.  Its steps include, first,
calculating the age-group specific trends using linear regression,
then, aggregation  using direct age standardization with fixed
weights. This method has been applied to risk factors only
\cite{dobsonst1,dobsonst2}, or to both, risk factors and
rates \cite{Kuulasmaa2000}. In the last case, the aggregated trends
were subject to correlation analysis in order to test MONICA
hypotheses. Some indication of problems in this approach come when the  time plots of the age-group and survey-specific
mean values of data items exhibit clear non-linearity  of time plots and diversity of these plots over age groups (see, for example,    \cite{surveydb}, POL-TARa, BMI).

A different modification uses the multiple logistic regression
procedure  applied to the whole set of data \cite{Edward2005}. As a
result, the marginal characteristics were obtained directly from the
procedure. This is equivalent to direct standardization, with
weights corresponding  to the analyzed population.

In examples above the method was applied to the samples having wide
age range (40 years and more) while spans  between consecutive
surveys were 3-10  years. Some studies of adolescents deal with
samples of age range  5-8 years, being  sampled every year or every
other year. In that case, the trends were first  examined visually by
age , since, as it was acknowledged, "they exhibit a wide diversity
of age-specific patterns" \cite {Kautiainen2002,Chen2003}.

One commonly used approach  to cope with such a diversity subdivides
the overall time period into several segments and the overall age
range into several age categories, for which the corresponding plots
suggest linear trends.  Alternatively,  trends are evaluated for
aggregated (age-standardized) parameters \cite{Kautiainen2002}.

Summing up, we may conclude that  methods used so far for the analysis of the changes of health-related indicators,  though being the best available ones at that time, suffered from one principal drawback: lack of causality. In turn, this is due to the fact that comparison is made between different entities, or, in terms of Dynamic Model, between different objects (see section \ref{prin}).

\section {Dynamic Modeling: History and Inventory of General Principles  \label{prin}}
In order to apply adequately the dynamic model paradigm to population health, we have to summarize the key notions and principles of this paradigm. We will start our inventory with the Newton's second law, which historically  is the very first dynamic model. We will end up this section with formulation of the main principles of Dynamic Modeling in form of definitions, labeled \textbf{P1} - \textbf{P4}.

\subsection {Newton's Second Law of Motion}
The following stuff is quoting from the English translation by Motte \cite{Newton}  of the Isaac Newton's formulation in Latin in  Principia, 1687.

"Law II (in English): The alteration of motion is ever proportional to the motive force impress'd; and is made in the direction of the right line in which that force is impress'd."

We would like  to use the structure of the above law's formulation  as golden standard for general definition. In this view,  the following comments are essential.

\begin{itemize}
\item The above law was expressed verbally. The mathematical expression for it, for example in modern form
\begin{equation}\label{eq:nfl}
m \frac{d \mathbf{v}}{dt}=\mathbf{F}
\end{equation}
is not fully equivalent to the original verbal form, since it does not specify what is cause and what is effect.
\item The law is not a kind of assumption to be used later in calculation, rather this is property of real world, derived by Newton from observation and logical inference.
\item The law postulates causality  clearly: the cause is  (motive) force, while the effect produced by this cause is  the alteration of motion.
\end{itemize}

\subsection {Definition of Dynamic Model in General Case}
Mathematical forms of  dynamic models, often being considered  as a complex of several interlinked dynamic models, are subject to special mathematical discipline - theory of dynamic (dynamical)systems (see, for example, \cite{Luenberger}). It is important to  stress, that our prime interest is in dynamic models, not in dynamic systems.

Usually, there were no problems in application of dynamic models in engineering and economy, as well as in interpretations of the results obtained.
The need for re-inventory of the notion of dynamic model, has been encountered, however,  when applying it in biology.

The following definition was given by \cite{Ellner} :
"Dynamic models are simplified
 representations of some real-world entity, in equations or computer
 code. They are intended to mimic some essential features of the
 study system while leaving out inessentials.
The models are called dynamic because they describe how system properties
 change over time."

We formulate the following definition:
\begin{description}
\item [\textbf{P1: }Definition of Dynamic Models:]
 Dynamic models are simplified
 representations of process of change (over time) of some real-world entity, in verbal and mathematical form.
 They are intended to mimic some essential features of the study object in process of change, while leaving out inessentials.
 The models are called dynamic because they describe changes of object properties over time, cased by driving forces.
( from Greek dynamikos "powerful", dynamis "power") ;
\end{description}
Thus, dynamic model operates with object, characterized by  set of properties, and Driving Force,
acting at object and causing changes of the object's essential properties.

Definition \textbf{P1}  differs from  one cited above in the following elements:
\begin{itemize}
\item \textbf{P1} defines that dynamic model represents process of change of real-world entity, rather than real-world entity itself.
\item In   \textbf{P1} "computer code" is excluded, since computer code may be an image of any process, possibly violating  physical laws. While we are concerned with  exploration of real world.
\item The models are "dynamic" not at all because of properties are changing. Rather because of the cause is postulated, generating these changes, -  Driving Force.
\end{itemize}
\begin{description}
\item [\textbf{P2: }Definition of Object:]
The real-world object is physical entity, traceable in time,
which means that it could be observed over some time interval, possibly small enough,
being virtually the same in common sense, and carrying the set of properties.
\end{description}
\begin{description}
\item [\textbf{P3: }Definition of State Variables:]An object is associated with a set of potentially measurable characteristics - State Variables.
These characteristics are assumed to be essential properties of the object. For each such variable, the rate of change is proportional
to the variable-specific net  driving force, acting upon the object.  Hence - State Variables are continuous and right-differentiable function of time.
\item [\textbf{P4: }Definition of Controls:]
The model (\ref{eq:nfl})  could be extended  by adding position vector and written in "Dynamic Systems" format:
\begin{eqnarray}\label{eq:vuz1}
\frac{d\mathbf{x_1}} {dt} &=& \mathbf{x_2} ,\quad
\frac{d\mathbf{x_2}} {dt} =  \frac{1}{m} \mathbf{u(t)} ,
\end{eqnarray}
where $\mathbf{x_1}$ is position vector,   $\mathbf{x_2}$ - velocity vector, $u(t)$ - represents driving force  $F$ , underlying the fact, that it could  be non-continuous function of time.

Note that in (\ref{eq:vuz1}) Driving Force for State Variable vector  $\mathbf{x_1}$  is State Variable vector   $\mathbf{x_2}$, thus being continuous, while Driving Force for State Variable vector    $\mathbf{x_2}$ is external force defined by  (vector) function $\mathbf{u(t)}$. We will use term "Controls" to call such an  external Driving Forces.
\end{description}
For a body moving  in gravitational field of a planet, State Variables position and velocity  "predict" further behavior of this  object, such as ability to "escape" the gravity of the planet without additional propulsion. In this respect State Variables differ from controls: the last ones modify State Variables rates of change, however they can not serve as predictors.

Using dot notation for differentiation  and combining vectors    $\mathbf{x_1}$  and   $\mathbf{x_2}$  into common vector $\mathbf{x}$,  equations  (\ref{eq:vuz1}) could be rewritten as:
\begin{equation}\label{eq:Fxu}
  \mathbf{ \dot{x} }= \mathcal{F}( \mathbf{ x , u(t) } )
\end{equation}
In verbal form, it postulates that rate of change of State Variables is a function of State Variables and controls.

For any real-world object, a dynamic model is built-up to facilitate the following three main practical tasks pertaining to this object:
\begin{description}
\item [\textbf{Task1:} Analysis:] Ultimately, design of function $\mathcal{F}$ is made by a researcher, dealing with a real-word object and its dynamics, depending on his/her intuition, experience and skill. The known prototypes play also important role. After design is made, the functional form is known up to a set of still unknown  parameters. The simplest case is linear form with coefficients to be set up. To evaluate these parameters, the measurements on State Variables and controls should be used, collected during some time period.  We will use term "Analysis" for this task.

\item [\textbf{Task2:} Prognosis:]If function  $\mathcal{F}$  is identified and State Variables are known for a moment $t_0$ and functions  $\mathbf{u(t)}$  are defined on interval $[t_o,T]$ then State Variables could be calculated for this interval, thus making prognosis of future behavior of the object.

\item [\textbf{Task3:} Control:] If criteria are set up, identifying the favorable future behavior of the object, and options are given for possible choice of controls as functions of time, then the task of control is to find out such a functions that optimize the given criteria.
\end{description}

\section {Dynamic Modeling in Health Research \label{ModPopLev}}

\subsection {Health Research, Individual Level}
In the following example we consider  a human-being  whose weight change over time is subject of some study. Let $\mathbf{x(t)}$ be a weight of the subject measured in standard way  at moment $t$. Due to measurements error it has a nature of random variable.  In addition,  the theoretical plot of this function over time will expose daily cycles, which are not of the study  primary interest. Rather we would like to operate with some smoothed characteristics of weight. We assume that according to study protocol, weight is measured every day at the same time in the morning, after "emptying body tanks"  before eating. So, we may think of sequence of measurement time moments $t_i$, where $i$ is sequence number of day since the beginning of the study.

We introduce function $v(t)$ defined as follows:

\begin{eqnarray}\label{eq:x2v}
v(t_i) & = &  E(x(t_i)), \quad t=t_i, \nonumber \\
v(t)   & = &  v(t_i) +  ( v(t_{i+1})- v(t_i)) \cdot
   \frac {t-t_i} {t_{i+1} - t_i }
, \quad  t_i \le t \le t_{i+1}
\end{eqnarray}

This function satisfies all  the requirements for the State Variable: it is continuous and right-differentiable function of time.
The current knowledge on  weight changes in adults could be summarized as follows.
Weight change in adult is, in fact, change in amount of body fat, which is determined by balance of calories taken with meal and burned throughout the  body  activity in over a certain time period. Thus, we can introduce function  $u(t)$ representing daily balance of calories, expressed in weight units
(see, for example, http://www.weightlossforall.com/calories-per-pound.htm
"One pound of body fat equals roughly 3,500 calories."). This function plays role of control for  $v(t)$ and we may postulate simple  model for weight change:
 \begin{equation}\label{eq:wevu}
   \dot{v} = u(t)
\end{equation}

We consider a  hypothetical study testing some technique  for weight reduction (it may include education, dietary recommendation, advice on physical activity etc.). Assume  that measurements of weight are available before and after the beginning of intervention, moment  $t_0$.

To highlight principal conceptual aspects, we make the following additional simplifying assumption:  $u(t)=u_1$, if $t\le t_0$,  $u(t)=u_2$, if $t> t_0$, where  $u_1$, $u_2$  are scalars.
In practice,  estimates for $u_1$ and $u_2$ could be obtained as slopes in linear regression models applied to measurements $x(t)$ for   $t\le t_0$ and  $t> t_0$ correspondingly.

 Condition   $u_2>u_1$  indicates that tested technique  is better than previous one (in practice, value $\alpha$ could be added having sense of "practical significance", so that condition will look like $ u_2>u_1 + \alpha $).
Note, that "success" is derived  from comparison of time trends for weight, not from the fact of decreasing weight for $t> t_0$.
Theoretically, positive $u_2$  may indicate success, if the weight growth has diminished, and negative  $u_2$  is not  a success, if  $u_1$ also was negative and approximately the same in value.

It is convenient to call all the data items, available in the study database and pertaining to a subject at certain moment of time,  measurements.

In dynamic model view, most of the measurements  fall into three categories:
\begin{description}
\item [\textbf{1: } State Variables: ] for example, age, weight, height, schooling years. Recall, that measurements for State Variables are not State Variables. They refer to each other as  $x(t)$ and $v(t)$ in the described above example.

\item [\textbf{2: } Modifiers:] for example, smoking status (smoking now, ex-smoker, or never-smoker), current physical activity, current dietary habits ( including 24 hours food consumption recall)
\item [\textbf{3: } Class indicators:] for example, sex, race, community, other characteristics, which are categorical and believed to be constant during the study time span.
\end{description}

Outside of the above categories are  multi-item outcomes of different questionnaires and tests. Some of them could result in one summary item ( for example, current physical activity level, which then could be classified as control). Questionnaire on smoking history may result in total amount of tobacco smoked so far. This indicator has a nature of State Variable. We leave further  consideration of this issue for future publications.

Similar to models in mechanics, Modifiers in health research modify status of body in terms of State Variables, however they can not serve as  predictors, for example, as predictor of instantaneous failure. Thus, current Hazard can not depend, for example,   on smoking or physical activity. This simple rule of dynamic model philosophy is widely violated in practice of methods used so far in health research.

Observe that continuous function of any number of State Variables is itself a State Variable,  and the same function applied to the measurements of corresponding variables serve as measurement for resulting variable. The expression (\ref{eq:wevu}) remains valid for this variable after control $u(t)$  is properly scaled.

In our future example we will deal with such a variable, The Body Mass Index,  defined as
\begin{equation}\label{eq:BMI}
    BMI=\frac{ weight(kg)} {height(m)^2}
\end{equation}
Here we pay tribute to tradition, using term weight instead of scholastically  more correct term "mass".

\subsection {Dynamic Model of Population: Heuristic Approach}

We may think of population as a collection of subjects identified for each calendar date by a certain rule. For example, for population of an urban  district such a rule may identify all permanent citizens having  home address within unambiguously defined  administrative boundaries. The rule must be the same throughout the calendar period for which  the population is supposed to be analyzed.

Health-related and other population characteristics, if available, has a form of age-distributed profiles, specific for calendar years, rather than individual-specific  measurements for all current subjects of population.

Measurements on population level are performed for random samples (stratified or not), taken, for example, every 5 year.

Thus, the challenge is how to adopt for population level  the dynamic model paradigm described so far for individual level.

To describe population history, it is convenient to use plane $(y,a)$, where  $y$  is real-valued calendar time in years, vertical axis,   $a$ is real-valued age in years, horizontal axis. Such a set up for axes anticipates further use of matrices  with indexes $y,a$, when the first index is row number (vertical coordinate). For consistent setup, we have to specify an observational frame  in terms of ranges $[y_{min},y_{max}]$ for $y$ and $[a_{min},a_{max}]$ for $a$.

Each subject may enter this population  due to birth (if $a_{min}=0$), or crossing left-low boundaries, or migration in.  Each subject may leave this population due to death, migration out or crossing the right-upper boundaries. If a subject with coordinates  $(y_0,a_0)$ is within the population during time $t$ ,  at that time it has coordinates $(y_0+t,a_0+t)$. Thus we may say, it is moving along cohort line.

Consider all subjects having coordinates on half-open interval $( (y_0,a_0-{\Delta}a), (y_0, a_0)]$ at time  $t=0$. At time  ${\Delta}t$ all those left in population will arrive at   $((y_0+{\Delta}t, a_0-{\Delta}a +{\Delta}t ), (y_0+{\Delta}t-{\Delta}a,a_0+{\Delta}t]$.  In other words, the birth cohort of width ${\Delta}a$  moves from $(y_0,a_0)$ to $(y_0+{\Delta}t,a_0+{\Delta}t)$. We may think of such a cohort as of a container moving on plane $(y,a)$. The contents of each container in process of movement is changed due to migration and death. If the rate of contents update is negligible (say, less than  1\% per year), we may ignore it in our analysis. If not, the analysis has to take this into account.

Each container fits the definition of the dynamic model object, if we regard the corresponding State Variable as  mean of State Variables for currently available subjects. The dynamic equation then could be obtained from ones for each subject, having form (\ref{eq:wevu}), by taking means of both sides:
\begin{equation}\label{eq:wevum}
   \dot{\overline{v}} = \overline{u}(t)
\end{equation}

Since the whole selected observational frame could be covered by collection of non-overlapping cohorts of selected width, we may conclude that, in case of population, the overall dynamic model is a collection of dynamic  models specific for each cohort.

\subsection {Dynamic Model of Population: Axiomatic Setup}

The theoretical abstraction for birth cohort is one of
infinitesimal  age range, characterized by multidimensional
distribution of  the parameters of interest, not by physical
subjects.

Let $\mathcal{C}$ be  a $2$-dimensional real compact:
$$
\mathcal{C}=\{(y,a):y \in [y_{min},y_{max}], a \in
[a_{min},a_{max}]\},
$$
where $y$ is calendar time in years and $a$ is age in years.

Consider a population defined on this compact, which suggests that
there potentially exists a set of random variables (r.v.)  $X_i$,
$i=1...k$ representing the corresponding set of measurable
indicators of interest (State Variables) defined at each point $(y,a)$ of compact
$\mathcal{C}$.
 In this paper we restrict ourselves to the case of
one indicator, so that subscript of $X$ will be omitted. To make the following description more illustrative let us keep in mind the Body Mass Index (BMI) as an example of the indicator in question.

We introduce the following notation
$$
v(y,a)= E(X(y,a)).
$$

For the sake of simplicity while describing the core dynamic model,
we assume,

\begin{equation} \label{eq:L301}
X(y,a)=v(y,a)+\epsilon \rm{, where }~E(\epsilon)=0,  ~ D(\epsilon)=
\sigma^2,~ \forall (y,a): (y,a)\in \mathcal{C}
\end{equation}

The dynamic equations describe changes of the distribution  of r.v.
$X$ for a birth cohort  taken at  point $(y,a)$ over time interval
$\ud t$:
\begin{equation}\label{eq:L302}
v(y+\ud t,a+ \ud t)=v(y,a)+u(y,a)\ud t+o(\ud t),  \rm{ where
}~\frac{o(\ud t)}{\ud t}  ~\rightarrow  ~0\rm{, as  } ~\ud t~
\rightarrow ~0.
\end{equation}

On one hand, function   $u (y,a)$  represents the rate and  direction of change of the State Variable  due to
the driving force generated by the environment. On the other hand, it is the driving force (control) itself, properly scaled.

The driving force at  $(y,a)$  does not depend on the properties
of the cohort passing at the time $y$ the age $a$.  Moreover,
theoretically, the very fact of its existence doesn't depend on
whether or not  there is a non-empty  cohort passing at the time $y$ the age $a$.

For the sake of convenience we will use terms "Mean levels" or
"levels" for the values of function $v(y,a)$, and "cohort trends" or
"C-trends" for the values of function $u (y,a)$.

 In the  advanced model  the function   $u (y,a)$  represents sum of  the environmental
force  and the force due to current state of the cohort. This will
lead to replacement of $u(y,a)$ in (\ref{eq:L302}) by
$u(y,a)+bv(y,a)$, where $b$ is a model parameter.

Let $v_0(y,a)$ be the value of $v(y,a)$ at low-left boundary of the
compact $\mathcal{C}$ for a (birth) cohort crossing the point
$(y,a)$:
\begin{equation}\label{eq:L303}
v_0(y,a) = v(y-\delta, a-\delta), \qquad \rm{where }~
\delta=min(y-y_{min},a-a_{min}).
\end{equation}

Then $v(y,a)$ can be expressed as
\begin{equation*}\label{eq:l304}
v(y,a)=v_0(y,a) + \int_0^\delta u(y-t,a-t)\ud t
\end{equation*}

Thus, if the values of  $v_0(y,a)$ at low-left boundary and $u(y,a)$
on $\mathcal{C}$ are known, then the function $v(y,a)$ could be
evaluated for each point on $\mathcal{C}$.

The generalization of the model (\ref{eq:L301}), (\ref{eq:L302}) for
the case of multidimensional distribution and state-dependent
dynamics is straightforward, by treating functions $v(y,a)$ and
$u(y,a)$ as vector functions, by replacing $D(\epsilon)= \sigma^2$
in (\ref{eq:L301}) by $Cov(\epsilon)=\Sigma$ and by replacement of
$u(y,a)$ in (\ref{eq:L302}) by $u(y,a)+bv(y,a)$, treating b as a
matrix.

\section{Dynamic Model of Population: Analytical Form \label{Anaform}}

\subsection {General Formulation of the Task }

Suppose that a set  of measurements is available  $(x_k,y_k,a_k)$,
$k=1,\ldots,K$, for subjects selected in a set of the independent
cross-sectional surveys. We assume that for each survey the
stratified by gender and age group random sample scheme was used.
The age group stratification could be different in different
surveys, however, for standard case, we assume that overall age
range is the same in all surveys.

The general formulation of the task is to estimate the functions
$v_0(y,a)$ and $u(y,a)$ on $\mathcal{C}$, using the available
measurements $(x_k,y_k,a_k)$, $k=1,\ldots,K$.

To solve this problem one option would be to formulate the
optimization problem in functional space:  to minimize  the
functional $I$:

\begin{equation}\label{eq:L305}
I(u,v_0) = \Bigg ( \sum \Big( x_k-v(y_k,a_k) \Big ) \Bigg )^2,
\end{equation}

applying some additional requirements on functions  $u(.,.)$ and
$v_0(.,.)$, such as   continuity (piece-wise continuity), and /or
restricted variation.

However, it seems more convenient  to transform the above problem
into the discrete - scale  analogue and to take the advantage of the
simplicity of  the analysis and  adaptation of the numerical methods
available in the standard statistical packages.

\subsection {Discrete-Scale Model}

Let $i$ and  $j$  be an integer value of  time  in years and an
integer value of age in years correspondingly. Our intention is to
build up the integer-values  proxies of the equations
($\ref{eq:L301}$ - $\ref{eq:L305}$).

Let $P(i,j)$  be a parallelogram-shaped element (convex hull)
defined by its angle points: $$  \{(i,j-1),~ (i,j),~ (i+1,j+1),~
(i+1,j)\}
$$ excluding its left and upper boundaries, which could be written
as
\begin{equation} \label{eq:B005}
P(i,j) \doteq \{(a,y): y \in [i,i+1),~ a \in ((j-1) +
(y-i),~j+(y-i)]\}
\end{equation}

We impose for function $u(.,.)$ the conditions of being constant on
each $P(i,j)$ and for functions  $v(.,.)$  being constant on $a$ and
linear on $y$ with constant  slope $u(i,j)$.

Formally this could be expressed as follows:

\begin{equation} \label{eq:B010}
u(y,a)=u(i,j), ~ \forall  i,j,y,a:~ (y,a) \in P(i,j)
\end{equation}

\begin{equation} \label{eq:B020}
v(y,a)=u(i,j) \cdot (y-i)+ v(i,j), ~\forall  i,j,y,a:~ (y,a) \in
P(i,j)
\end{equation}

We derive minimal and maximal values for $i$ and $j$ from the
correspondent values for $y$ and $a$ using definition
($\ref{eq:B005}$):
\begin{eqnarray}\label{eq:B301}
(i_{min},j_{min}): (y_{min},a_{min}) \in  P(i_{min},j_{min}), \\
(i_{max},j_{max}): (y_{max},a_{max}) \in
P(i_{max},j_{max})\nonumber
\end{eqnarray}

For convenience,  from now on we will use relative scale for age and
time, defined by transformation
$$
i - i_{min} \rightarrow i\rm{, }~ j - j_{min} \rightarrow j
$$

Consider functions $u(i,j)$ and $v(i,j)$ defined on integer-valued
two dimensional domains
\begin{eqnarray}\label{eq:uvdef}
\mathcal{U}&=&\{(i,j):i \in [0,I],~ j \in [0,J]\}, \nonumber\\
\mathcal{V} &=& \{(i,j):i \in [0,I+1],~j \in [0,J+1]\},
\end{eqnarray}

correspondingly, where
$$
 I=i_{max} -i_{min}\rm{, }~ J=j_{max} -j_{min}
$$

Now the main dynamic equation ($\ref{eq:L302} $ ) could be rewritten
as

\begin{equation}\label{eq:B302}
v(i+1,j+1)=v(i,j)+u(i,j),  ~\forall (i,j)\in  \mathcal{U}
\end{equation}

Let $v_0(i,j)$ be the value of $v(.,.)$ at low-left boundary of the
domain $\mathcal{V}$ corresponding to a (birth) cohort crossing the
point $(i,j)$:
\begin{equation}\label{eq:B303}
v_0(i,j) = v(i-\delta, j-\delta), \rm{where }¨~  \delta=min(i,j).
\end{equation}

Combining ($\ref{eq:B302}$) and  ($\ref{eq:B303}$), we rewrite
equation (\ref{eq:L303})  as:

\begin{equation}\label{eq:B304}
v(i,j) =v_0(i,j) + \sum_{m=1}^\delta u(i-m,j-m)
\end{equation}

From ($\ref{eq:B304}$) it follows that if $v(i,j)$ is set up on the
low-left boundary of  $\mathcal{V}$ and $u(i,j)$ is set up on the
whole $\mathcal{U}$ then $v(i,j)$ could be calculated for the whole
$\mathcal{V}$.

Finally, assembling ($\ref{eq:L301}$), $\ref{eq:B304}$) and
($\ref{eq:B020}$) for each available observation  $(x_k,y_k,a_k)$,
$k=1,\ldots,K$, we obtain:

\begin{eqnarray}\label{eq:B305}
x_k =v_0(i,j) + \sum_{m=1}^\delta u(i-m,j-m) + (y_k-i)\cdot u(i,j)
+\epsilon_k \rm{,}
 {}\nonumber\\
\rm{ where }~Var(\epsilon_k)=\sigma^2, ~
Cov(\epsilon_k,\epsilon_l)=0,~if~ k \neq l
\end{eqnarray}

Let $\mathbf{z}$  be a vector  with components $v_0(i,j)$ and
$u(i,j)$ ordered in the following way:

\begin{eqnarray}\label{eq:vuz}
  \mathbf{v}_0 &=& \big(v(I+1,0),\ldots,v(0,0), \ldots,
  v(0,J+1)\big)^T \nonumber\\
  \mathbf{u} &=& \big(u(0,0),\ldots,u(0,J),\ldots,
  u(I,0),\ldots,u(I,J) \big)^T \nonumber\\
  \mathbf{z} &=& \big( \mathbf{v}_0^T \quad \vline \quad
  \mathbf{u}^T\big)^T
\end{eqnarray}

Using vector $\mathbf{z}$  and introducing vector of coefficients
$\mathbf{b}_k$, we  can rewrite ($\ref{eq:B305}$) in the form
\begin{equation}\label{eq:B306}
x_k =(\mathbf{b}_k, \mathbf{z}) +\epsilon_k, ~\rm{where }
~Var(\epsilon_k)=\sigma^2, ~ Cov(\epsilon_k,\epsilon_l)=0,~if~ k
\neq l
\end{equation}

This form represents a particular case of Gauss-Markov Setup for the
Least Squares Linear Estimation problem \cite{Rao}.

Let $\mathbf{B_0}$ be a matrix composed of row vectors
$\mathbf{b}_k^T $ in  ($\ref{eq:B306}$), $\mathbf{z} $ and
$\mathbf{x}_0 $ stand for column vectors of the parameters
$\mathnormal{z}_j $ and the variables $\mathnormal{x}_k $
correspondingly, and $\mathnormal{S}_0 $  be a scalar function
defined as
\begin{equation*}\label{eq:B307}
\mathit{S}_0( \mathbf{z} ) =( \mathbf{B_0}\mathbf{z} -
\mathbf{x}_0)^T  ( \mathbf{B_0}\mathbf{z} -  \mathbf{x}_0)
\end{equation*}

Note that if $ \mathit{rank}(\mathbf{B_0} ) = \mathit{dim}(
\mathbf{z} ) $, then  estimates obtained by unconditional minimizing
of function $\mathnormal{S}_0 ( \mathbf{z} ) $  are unique ones.
Such a case takes place only if the observations cover all the
elements   $P(i,j)$  when  surveys cover the whole analysis period
without gaps.

In practical cases, minimizing of  $\mathnormal{S}_0  $ results in
singular or ill-posed Inverse Problem, and  so-called regularization
techniques  are needed to obtain meaningful solution estimates. Most
of these techniques employ the idea of smoothing of some function
having clear physical interpretation \cite{Neumaier}.

Here we suggest one such  technique for smoothing.
\subsection{Smoothing}

We define the following  indicator of smoothness of function
$v(.,.)$
\begin{eqnarray}\label{eq:B401}
\mathnormal{S}_1( \mathbf{z} ) & =  \sum_{i=0}^{I+1} \sum_{j=1}^{J}
\Big( v(i,j-1)-2v(i,j)+v(i,j+1) \Big)^2
+ {}\nonumber\\
 &  \sum_{j=0}^{J+1} \sum_{i=1}^{I}    \Big( v(i-1,j)-2v(i,j)+v(i+1,j)  \Big)^2
\end{eqnarray}
Each term in this sum represents the  square for a proxy  of the
second derivative of function $v(.,.)$  with respect to age or with
respect to calendar time at point $(i,j)$.

Replacing $v(.,.)$ by $v_0(.,.)$ and $u(.,.)$ using
($\ref{eq:B304}$), and the last ones by vector  $\mathbf{z}$, we
will transform the previous expression to the following form:
\begin{equation}\label{eq:B402}
\mathit{S}_1( \mathbf{z} ) =( \mathbf{B_1}\mathbf{z}- 0 )^T  (
 \mathbf{B_1}\mathbf{z} - 0 )
\end{equation}
Similarly, we define indicator of smoothness of function  $u(.,.)$
\begin{eqnarray*}
\mathnormal{S}_2( \mathbf{z} ) & =  \sum_{i=0}^{I} \sum_{j=1}^{J-1}
\Big( u(i,j-1)-2u(i,j)+u(i,j+1) \Big)^2
+ {}\nonumber\\
 &  \sum_{j=0}^{J} \sum_{i=1}^{I-1}    \Big( u(i-1,j)-2u(i,j)+u(i+1,j)  \Big)^2
\end{eqnarray*}
allowing form
\begin{equation}\label{eq:B4021}
\mathit{S}_2( \mathbf{z} ) =( \mathbf{B_2}\mathbf{z}- 0 )^T  (
 \mathbf{B_2}\mathbf{z} - 0 )
\end{equation}
Now we can add one or both constraints  ${\mathit{S}_k( \mathbf{z} )
\leq \alpha_k }$ with some selected  ${\alpha_k \geq 0}$,  $k=1,2$,
to the model (${\ref{eq:B306}}$). Observe that indicators
$\mathit{S}_0$, $\mathit{S}_1$, $\mathit{S}_2$ are quadratic
functions in finite vector space $ \mathrm{E}_n $ with elements
(vectors) $\mathbf{z}$ and $n=dim( \mathbf{z} )$. The optimization
problem for point estimation for our case,  could be formulated as
\begin{equation}\label{eq:P1}
\min_{\mathbf{x} \in \rm{E}_n }  S_0( \mathbf{x} ) \textrm{, subject
to } {S_k( \mathbf{x} ) \leq \alpha_k } \textrm{, with given
}{\alpha_k
>0 }, ~k=1,2.
\end{equation}
Let $n_0$ , $n_1$ and $n_2$  be numbers of rows in matrices
$\mathbf{B}_0$ , $\mathbf{B}_1$ and $\mathbf{B}_2$ correspondingly. Let
$\lambda_1$,  $\lambda_2$  be some non-negative  scalars.
Introducing matrices and vectors
\begin{equation}\label{eq:p11}
\mathbf{B}= \left( \begin{array}{c}
\mathbf{B}_0 \\
\hline
\mathbf{B}_1 \\
\hline
\mathbf{B}_2 \\
\end{array} \right),
\quad \mathbf{x}= \left( \begin{array}{c}
\mathbf{x}_0 \\
\hline
\mathbf{0}\\
\hline
\mathbf{0}\\
\end{array} \right),
\quad \mathbf{W}= \left( \begin{array}{c|c|c}
\mathbf{I}_0 & 0 & 0 \\
\hline
0 & \lambda_1 \mathbf{I}_1 & 0\\
\hline
0 & 0 & \lambda_2 \mathbf{I}_2 \\
\end{array} \right)
\end{equation}
where $\mathbf{I}_0$, $\mathbf{I}_1$ and  $\mathbf{I}_2$ are identity matrices of rank  $n_0$ , $n_1$ and $n_2$ correspondingly, we
can formulate the problem of least squares estimation in the
following form (a modification of Gauss-Markov setup which fits form
of Aitken setup \cite{Rao}
\begin{equation}\label{eq:p12}
\mathbf{x}=\mathbf{B}\mathbf{z}+\mathbf{\epsilon},\quad
E(\mathbf{\epsilon})=\mathbf{0}, \quad
D(\mathbf{\epsilon})=\sigma^2\mathbf{W}^{-1}
\end{equation}
for which the point estimation problem is
\begin{equation}\label{eq:p13}
\min_{\mathbf{z} \in \mathrm{E}_n }  S( \mathbf{z} ) \textrm{,
where} S( \mathbf{z} ) =( \mathbf{B}\mathbf{z} - \mathbf{x})^T
\mathbf{W} (\mathbf{B}\mathbf{z} -  \mathbf{x})= S_0
(\mathbf{z})+\lambda_1 S_1(\mathbf{z})+\lambda_2 S_2(\mathbf{z})
\end{equation}
\cite {MolMic} have shown that problems (\ref{eq:P1}) and (\ref{eq:p13}) are equivalent:  problem  (\ref{eq:P1}) with given
$\alpha_1$, $\alpha_2$  possesses  the same solution as problem  (\ref{eq:p13}) with some $\lambda_1$, $\lambda_2$, and vice versa, or both don't possess any solution.

Since part of its components are set to zero,  the data vector $\mathbf{x}$ in
(\ref{eq:p11}) could not be treated as a "true" data vector if the
problem is considered from the classical frequentist prospective.
The last one is based on retrospective evaluation of the procedure
used to estimate  parameters over the distribution of possible data
values conditional on the true unknown values of parameters
\cite{Gel1}, p.7. The logically consistent treatment of the problem
is based on Bayesian paradigm, where statistical conclusions about
unknown parameters are made in terms of probability statements,
conditional on observed data. As noted in \cite{Gel1}, p.7, in
despite this difference , it will be seen that in many simple
analyses, superficially similar conclusions result from the two
approaches to statistical inference. In particular, this concerns
Bayesian analysis of the classical regression model: under a
standard noninformative prior distribution, the Bayesian estimates
and standard errors coincide with the classical results \cite{Gel1},
p.235. The last statement justifies use of classical formulas and
numerical procedures for our "non-classical" case.

The question of primary practical importance is the existence of a unique solution for the problem (\ref{eq:p13}).

The following statement is proofed in \cite {MolMic}:
\paragraph{Corollary 1}
For  existence of a unique  solution to problem (\ref{eq:p13}) it
is sufficient to have 4 data points such that the corresponding
points $(y,a)$ on plane $y,a$ satisfy condition: no any 3 of
them are located on a common straight line.


\subsection{Outlines of the Algorithms. Setting up the Regularization Parameters}
As soon as parameters $\lambda_1$,  $\lambda_2$  are given in
setup (\ref{eq:p11}, \ref{eq:p12}), the following could be obtained routinely:
$\mathbf{\hat{z}}$  - point estimate of vector $\mathbf{z}$,
covariance matrix of this estimate $Cov(\hat{\mathbf{z} })$, and $\hat{\sigma^2}$ - estimate of ${\sigma^2}$.

Using functions  $u(i,j)$ and $v(i,j)$  defined  in (\ref{eq:uvdef}), we can create matrices

$\mathbf{V}: v_{i,j}=v(i+1,j+1)$,

$\mathbf{U}: u_{i,j}=u(i+1,j+1)$,

and vectors

$\mathbf{v}=(Shape(\mathbf{V},1))^T$,

$\mathbf{u}=(Shape(\mathbf{U},1))^T$,

where Shape is matrix function reshaping the original matrix into resulting one with different number of rows and columns (available, for example, in SAS/IML  \cite{SAS9.1IML}). In our case, results are vectors with consequently concatenated rows of the original matrices.

Each element $v_{i,j}$ of matrix  $\mathbf{V}$ corresponds to element $v_k$ of vector  $\mathbf{v}$ with
\begin{equation}\label{eq:ij2k}
k=(i-1)\cdot ncol( \mathbf{V} )+j,
\end{equation}
where  ncol(.) is matrix function returning number of columns.
Similar rule could be applied for linking  $\mathbf{U}$ and  $\mathbf{u}$.

Definition of vector $\mathbf{u}$    in (\ref{eq:vuz}) and expression for functions $u(i,j)$ in  (\ref{eq:B304}) allow to construct matrices

$\mathbf{A_{z2v}}: \quad \mathbf{\hat{v}}=\mathbf{A_{z2v}} \mathbf{\hat{z}}$, and

$\mathbf{A_{z2u}}: \quad \mathbf{\hat{u}}=\mathbf{A_{z2u}} \mathbf{\hat{z}}$.

Hence, the covariance matrices could be derived as

$Cov (\mathbf{\hat{u}})= \mathbf{A_{z2u}} Cov(\hat{\mathbf{z} }) \mathbf{A_{z2u}}^T$,

$Cov (\mathbf{\hat{v}})= \mathbf{A_{z2v}} Cov(\hat{\mathbf{z} }) \mathbf{A_{z2v}}^T$,

from which the corresponding matrices of Pearson's correlation coefficients  $\mathbf{R_v}$ and  $\mathbf{R_u}$ could be routinely produced.

Consider two consecutive level estimates
$\hat{v}(i,j)$, $\hat{v}(i,j+1)$  allocated along age axe
(similar consideration could be applied to allocation along calendar
years.

The coefficient of correlation for these estimates could be derived
from $\mathbf{R_v}$ applying rule  (\ref{eq:ij2k}). Let denote it $r_{a,i,j}$. Similarly, coefficients of correlation  $r_{y,i,j}$ could be derived for estimates $\hat{v}(i,j)$, $\hat{v}(i+1,j)$  allocated along years axe.

Consider task of predicting estimate $\hat{v}(i,j+1)$ using linear
predictor based on  $\hat{v}(i,j)$. The expression
$1- r_{a,i,j}^2$ is proportion of  "unexplained" part of variance of  $\hat{v}(i,j+1)$,
(see, for example, \cite{Rao} p.266). This part could be interpreted as "new information", or "signal", while
 $r_{a,i,j}^2$  could be regarded as proportion of "Noise".
The better smoothness is associated with lower signal-to-noise  ratio.
We combine all local indicators of smoothness into one common vector
\begin{equation}\label{eq:smu}
\mathbf{f_{v}} = Shape(\mathbf{V_{sma}},1) \|  Shape(\mathbf{V_{smy}},1), \quad where \quad
 v_{sma,i,j} = 1- r_{a,i,j}^2,  \quad
 v_{smy,i,j} = 1- r_{y,i,j}^2
\end{equation}
Vector $\mathbf{f_{u}}$ could be defined in similar way.

For practical use we have to select function, producing sample statistics for a vector-argument, such as mean,
median, minimum or a value of one predefined component and a target value for this statistics, $f_{sm}$. Let $fstat$ be
generic name for such a function. Then iterations are run by selecting $\lambda_1$  and   $\lambda_2$ until the following condition is satisfied
\begin{equation}\label{eq:smdelta}
max( abs( log( fstat_v(\mathbf{f_{v}} ) )-log(f_{smv}) ),
   abs( log( fstat_u (\mathbf{f_{u}} ) )-log(f_{smu}) )
   \leq \delta,
\end{equation}
where $\delta$  is  predefined accuracy.

With increasing values of $f_{smv}$, $f_{smu}$ the corresponding lines and
surfaces visually become smoother. For level estimates, for example,  if
$f_{smv} \rightarrow 0$, then   $\lambda_1 \rightarrow\infty$, and
solution converges to 4-parametric surface (\cite{MolMic}).

To measure difference in C-trends over age and calendar year, the pairwise comparison tests are performed for
mean values of C-trends, evaluated  for a set of age-year clusters, defined  by cluster sizes, $\Delta_a$ and $\Delta_y$.

Let  $\mathbf{U_{c}}$ be matrix of such mean values,   $\mathbf{u_{c}}$=$Shape(\mathbf{U_{c}},1)^T$ and matrix  $\mathbf{A_{u2uc}}$ such that  $\mathbf{u_{c}}= \mathbf{A_{u2uc}} \mathbf{\hat{u}}$. As soon as, matrix $\mathbf{A_{u2uc}}$ is created for given  $\Delta_a$, $\Delta_y$, $\mathbf{U_{c}}$  could be calculated, as well as  variance/covariance values for its elements in a format of

$\mathbf{C}=  Cov(\mathbf{\hat{u_c}})= \mathbf{A_{u2uc}} Cov(\hat{\mathbf{u} }) \mathbf{A_{u2uc}}^T$.

For each cluster, statistics and corresponding probabilities are
computed for  pairwise comparison of mean C-trends for current cluster
and  for adjacent one for older age group, and for current one and
for adjacent one for the next calendar years period ( if the
corresponding clusters exist). Using classical paradigm, this is done by testing linear
hypotheses in form

$\mathbf{H}_0: {u_c}_i-{u_c}_j=0$.

General expression for F-value (see for,
example, SAS/Stat manual, \cite{SAS9.1STAT}) in this case takes a simple form
\begin{equation*}\label{fexpr}
F = \frac   {({u_c}_i  -   {u_c}_j)^2 } { c_{i,i}- 2 c_{i,j}+ c_{j,j}}
\end{equation*}
Corresponding probability is computed using SAS function $probF$ (see
\cite{SAS9.3L}) as

$ Pr = 1- probF(F,1,n-r)$

Note, that in Bayesian view, these probabilities should be referred to as tail-area probabilities for posterior predictive distributions (\cite {Gel1} , p.169).

The algorithm, implementing the above outlines,  is written in SAS code using SAS products
(\cite{SAS9.3L},  cite{SAS9.2PROC}, \cite{SAS9.1IML}, \cite{SAS9.1STAT}  \cite{SAS9.1GRAPH}).
Results of pairwise tests are presented graphically in figure, produced  by PROC GCONTOUR, properly annotated ( see Figure \ref{Chart} in example of application).

For reference, we will call this algorithm DRM2(R), with prefix DRM2 to differentiate it from those developed in \cite{MolMic}.
We have built up also the modification of this algorithm, processing aggregated data, DRM2(A), thus DRM2(R) for "oRiginal" data, and
DRM2(A) - for "Aggregated" data.

Original individual data may be of quite big size, which reflects row number $n_0$ of matrix $\mathbf{B}_0$ in (\ref{eq:p11}), and hence, the required memory and time for calculation.

Aggregation is applied to original measurements  $(x_k,y_k,a_k)$, $k=1,\ldots,K$, producing summary statistics for $(age \cdot year)$ cells with size 1. As a result, arithmetic means are produced  $(\overline{x}_c , \overline{y}_c)$, number of original measurements  in each cell  $(n_c)$, and  ${S_{CSS}}_c$ -Corrected Sum of Squares, where $c=1,\ldots,C$ - collection of non-empty cells.
Matrix  $\mathbf{B_0}$ and vector $\mathbf{x}_0$  in (\ref{eq:p11} ) should be replaced by
$\mathbf{\overline{B}_0}$ and $\mathbf{\overline{x}}_0$, with cell-specific rows.

Let $\mathbf{n}_0$ be a frequency vector with components  $(n_c)$. The following expressions are essential elements of the DRM2(R) algorithm.

Contribution to cross-products $B \cdot x$ and $B\cdot B$:

EXPR1: $  \quad (\mathbf{\overline{B}_0}^T \cdot Diag(\mathbf{n}_0 ) \cdot \mathbf{\overline{x}}_0) $

EXPR2: $  \quad (\mathbf{\overline{B}_0}^T \cdot Diag(\mathbf{n}_0 ) \cdot \mathbf{\overline{x}}_0) $

Contribution to sum of squares of error terms,

$
EXPR3: \quad
(\mathbf{\overline{B}_0} \mathbf{\hat{z}} - \mathbf{\overline{x}}_0 )^T  \cdot Diag(\mathbf{n}_0 )  \cdot
(\mathbf{\overline{B}_0} \mathbf{\hat{z}} - \mathbf{\overline{x}}_0 )
+ \sum_{c=1}^{C}  {S_{CSS}}_c
$

Note, that DRM2(A) and DRM2(R) will produce identical outputs if all   $y_k$  within cells are equal.

\section{Example of Application \label{Exa}}
\subsection{Data}
To illustrate the method and to demonstrate its performance, the
data  from three cross-sectional surveys, conducted in North Karelia,
Finland,  during the period  1982 -1992, will be used. Formally this
set of data can be characterized as follows:
\begin{itemize}
\item Study population:  North Karelia, Finland.
\item  Study period: 1982-1992;
\item  Source of data: cross sectional independent surveys conducted in
years 1982, 1987, 1992
\item Sampling frame for each survey: the stratified by 10-year age groups
(25-34, 35-44, 45-54 and 55-64) and gender random sample.
\end{itemize}
The following  specifications defines sub-sample of  records and
items selected for analysis.

Only data for men will be used; the number of examined men in years
1982, 1987, 1992  is equal to  1537, 1481 and 673, correspondingly.

Original measurements of interest are: gender, date of
birth, date of examination, weight and  height.

The analysis variables included in the model:

BMI - the Body Mass Index, defined as $ {weight (kg)} / {height
(m)}^{2} $.

AGE - age in full years, defined as  year of examination  minus year
of birth.

YEAR - date of examination measured in years.

All surveys have started at  the beginning of the year, surveys 1982
and 1987 have   been completed in 4 months, survey 1992 - in three
months period.
\subsection{Analysis Setup}
The algorithm modification DRM2(A) have been used, preprocessing original data into aggregated format. There were 120 aggregated observations, 40 for each survey year.
The analysis was set up for the age range 25-64 and for the calendar
year period 1982-1992. In rule, controlling iterations, (\ref{eq:smdelta}), smoothing factors $f_{smv}$, $f_{smu}$  were set to 0.2, accuracy level, $\delta$ was set to 0.05; cluster sizes, $\Delta_a$, $\Delta_y$,  for producing comparison tests were set to 5.

We have found that for practical purposes it is enough to use
one selected point, $(1,1)$  for $fstat_v$,  and  one selected point, $(int( ncol( \mathbf{U})/2), int(nrow( \mathbf{U}) /2))$, for $fstat_u$.
\subsection{Results}
The results of the analysis are visualized by the set of
3-dimensional figures.

\begin{figure}
\begin{center}
\centerline{\includegraphics[angle=-90,width=20.5cm]{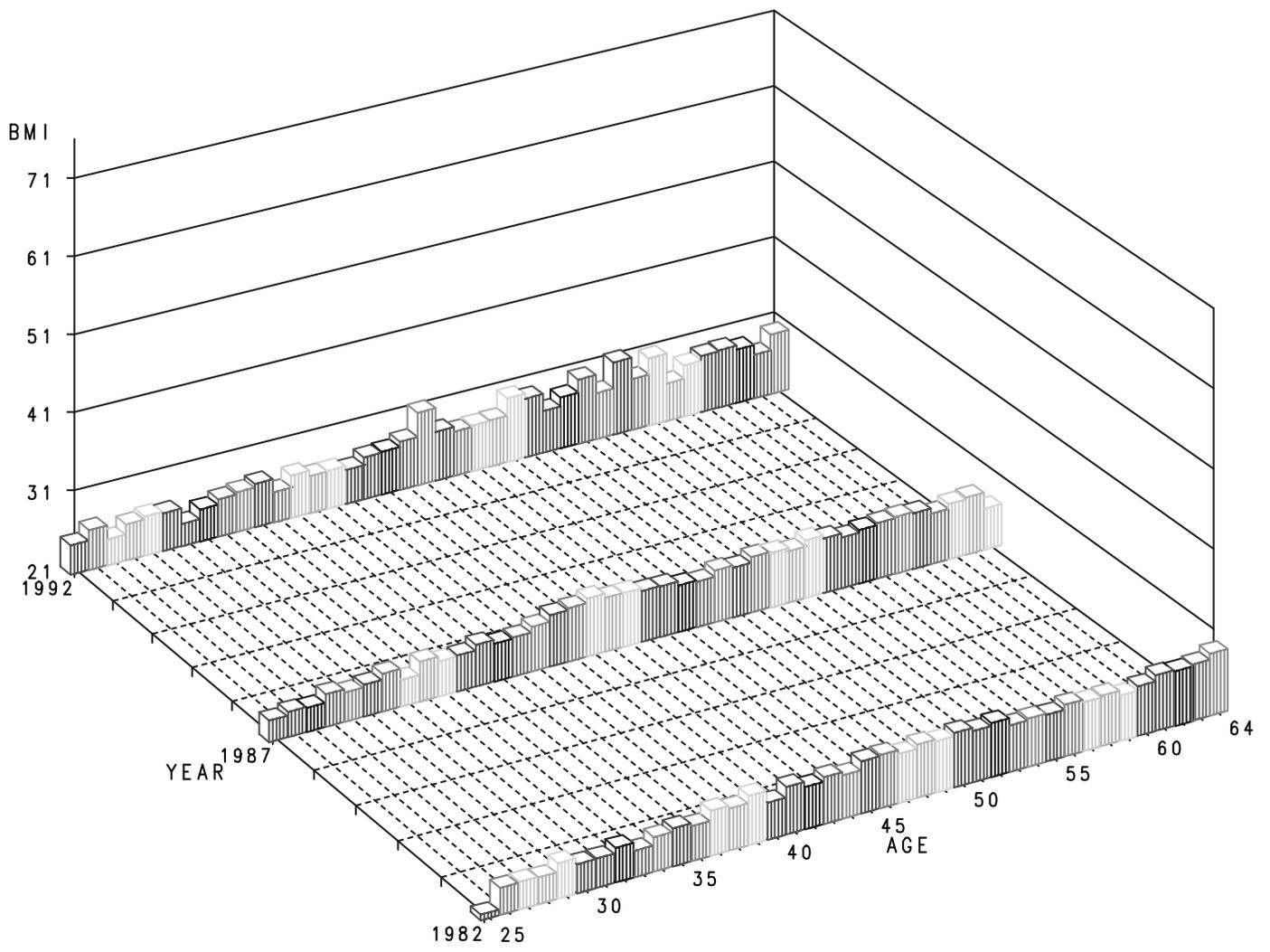}}
\end{center}
\caption{BMI, Men, Survey data. Means by year and age.
\label{ori}}
\end{figure}

Figure ~\ref{ori} displays the values representing  means of BMI
calculated for each age and year, for which the survey data are
available (number of cases in each cell exceeds 9). To visualize the
along-cohort changes, the columns corresponding to the same birth
cohorts in different surveys have similar shades of grey.

\begin{figure}
\begin{center}
\centerline{\includegraphics[angle=-90,width=20.5cm]{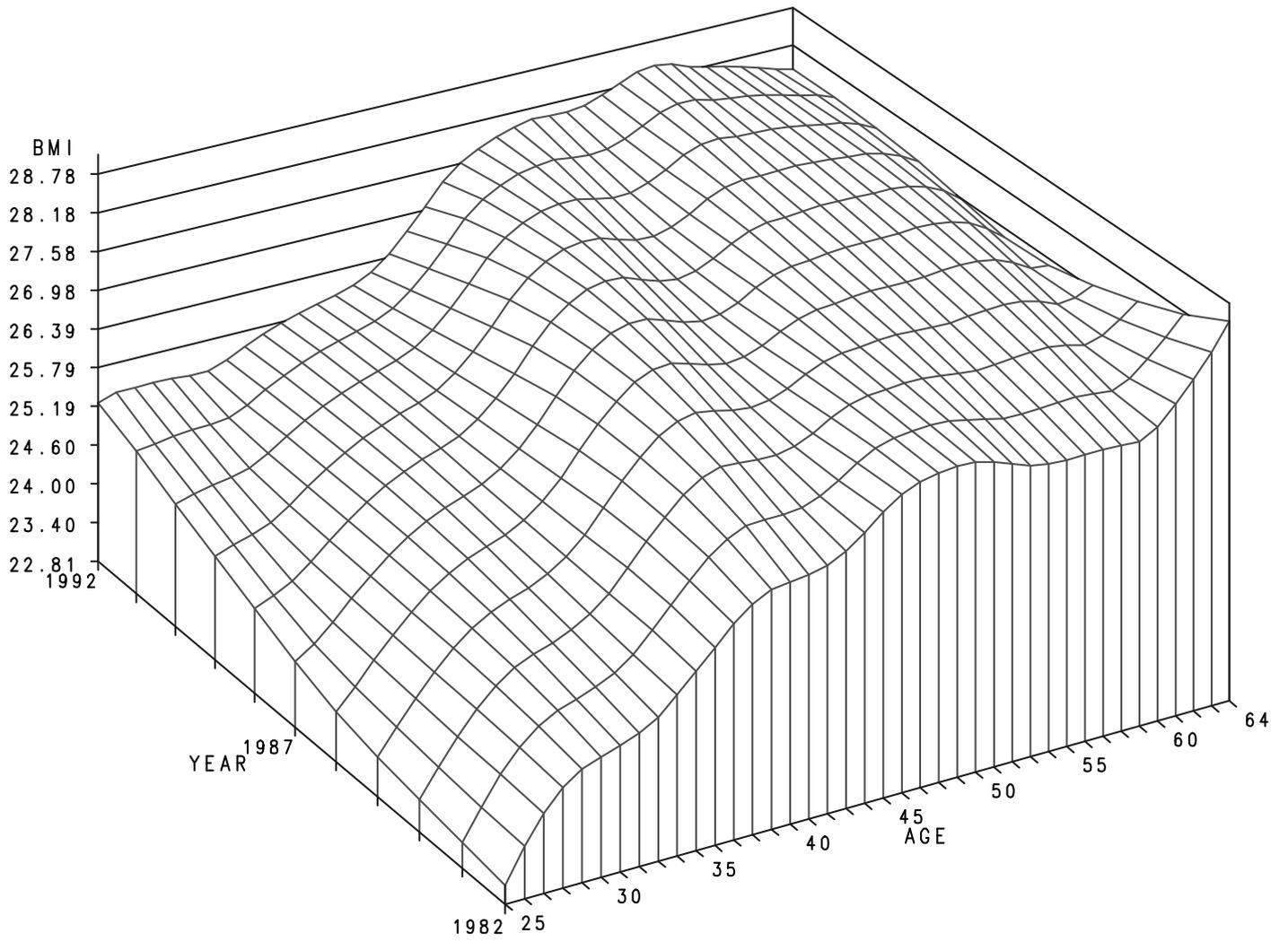}}
\end{center}
\caption{BMI, Men, Estimates of means by year and age.
\label{v}}
\end{figure}

Figure~\ref{v} displays estimates for the mean levels of BMI for the whole
domain, with study age range plus one year, and study period plus one
year.

\begin{figure}
\begin{center}
\centerline{\includegraphics[angle=-90,width=20.5cm]{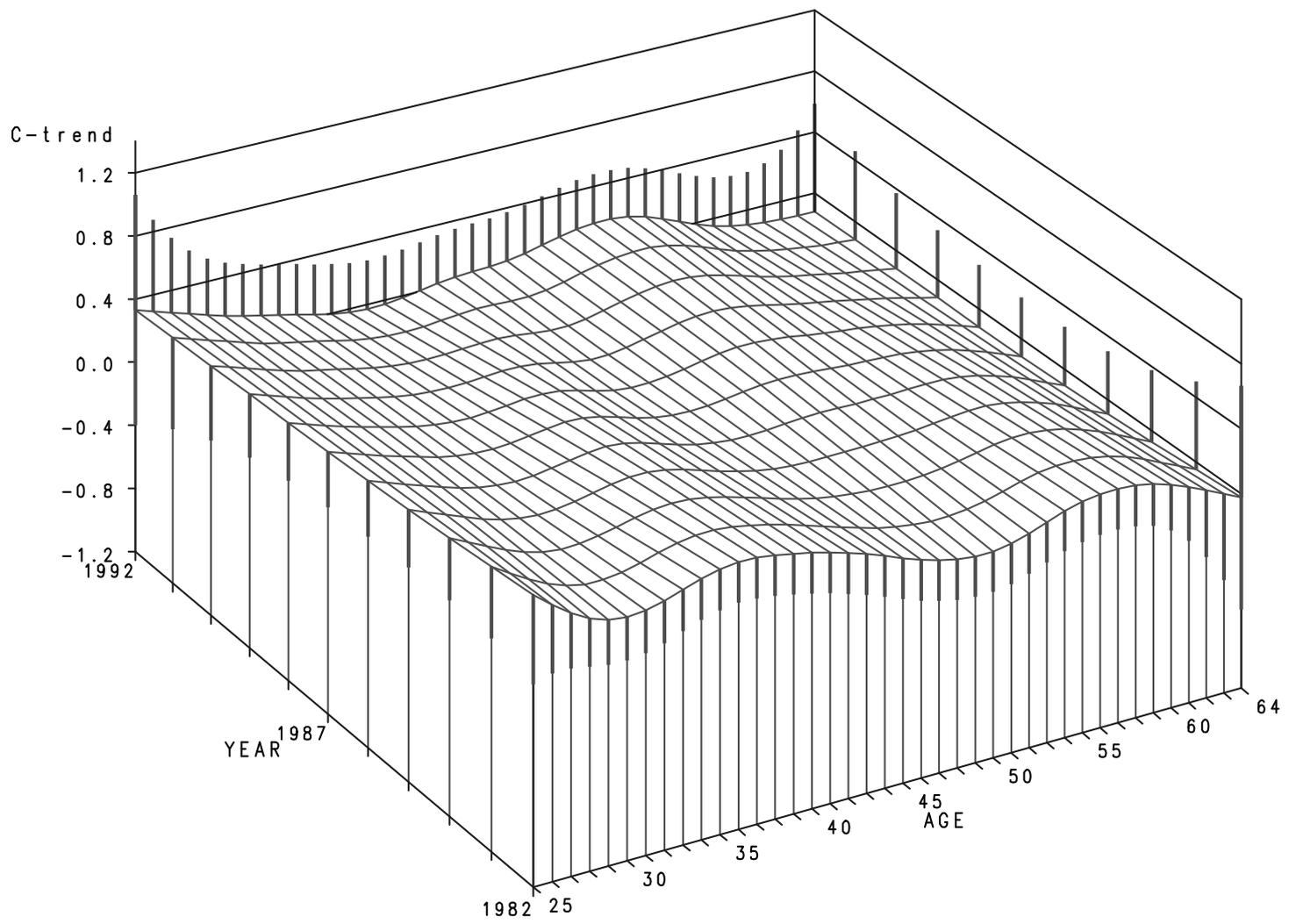}}
\end{center}
\caption{BMI, Men, Estimates of C-trends by year and age.
\label{u}}
\end{figure}

Figure ~\ref{u} displays C-trends with 95\% confidence intervals, shown at
left and front boundaries only.

\begin{figure}
\begin{center}
\centerline{\includegraphics[angle=-90,width=20.5cm]{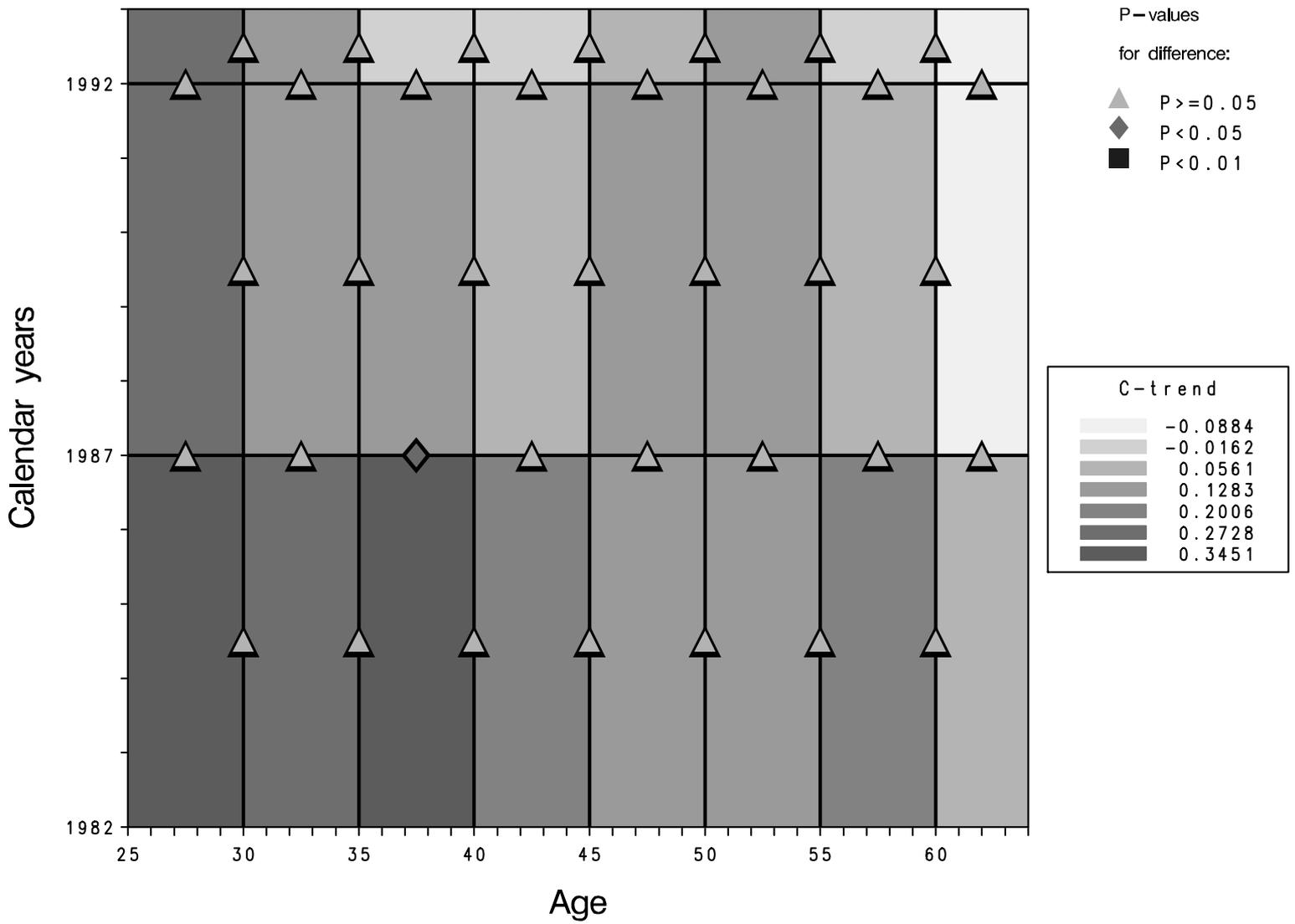}}
\end{center}
\caption{BMI, Comparison of C-trends  by clusters of age and calendar years.
\label{Chart}}
\end{figure}

Figure ~\ref{Chart} displays mean levels of C-trends  for specified age-year clusters, with P-values for differences between clusters.

These figures illustrate the principle "one figure is better than
one hundred tables", though all the underlying data are available
and could be presented in a set of tables.

Figure~\ref{v} shows that mean BMI levels increase along cohort lines
throughout the study period, although they are different for
different birth cohort. Specific peaks and troughs   follow cohort
lines.

Recall that C-trends represent the net external Driving Force (Modifier) causing changes over
time in cohorts. Therefore,
changes in C-trends  pattern over calendar years may indicate effect of
preventive activities, while difference across age range may
indicate both, age-specific uncontrolled changes and/or different
susceptibility to prevention.

In our case, Figure ~\ref{Chart}  shows  clear decrease of C-trends  in the period 1987-1992 compared with
the period 1982-1986 in the age range   35-40 $(p < 0.05)$;
No other significant differences between adjacent clusters were detected.

The further detailed analysis and final interpretation of the results
may require a log of the events
affecting the socio-economic and health care profiles of the study
population during the  study period. For example, a feasible
explanation of the observed effect in C-trends in age range 35-40
could be associated with creating new working places in years
1987-1992, which have decreased population flow out of the area,
taking place in years 1982-1986 in this age range and leading to
negative health selection (subjects with low BMI were leaving the area in searching for job places).

Summing up, we can conclude  that, in general, clustering of C-trends
looks reasonable, so we can use the results of
comparing  C-trends levels in adjacent clusters for our analysis.

\section{Conclusion and Discussion \label{conclu}}
In this paper we have presented  a novel formulation of  the key principles of dynamic modeling in general, and in application to health research, which justify the structure and interpretation of the core models dealing with  C-trends. In particular,
according to these principles, traditional  risk  factors' indicators  fall into
two categories, State Variables and Modifiers (see section \ref{prin} ),  having  different dynamical nature and, hence, playing different roles in the model and analysis.

As corollary of this, circular trends for State Variables have no sense at all. At the same time, only State Variables may determine instantaneous hazard rate of failure.
In dynamic models,  causality is postulated: changes are due to Driving Forces (Modifiers), existing in the real world.
In case of consecutive survey data, C-trends are believed to be proxies for Driving Forces,  providing the  tool
for three main practical tasks: analysis, prediction and control of  health on population level ( see section \ref{prin})

We have used these principles as a framework for developing  the dynamic model  of simulating the temporal changes in characteristics of a real-world
object - population. In the course of this process, first, we
have identified two interacting objects, population and its
environment, on the top aggregation level. Further system analysis
has led us to breaking down the study population  into a set of
potentially infinitesimally narrow birth cohorts, carrying over time
health state profiles expressed in terms of health related
indicators (State Variables).

The model employs the \emph{health field} concept, suggesting
existence of  an influencing factors (Modifiers), generated by environment and acting
on the population, specific for each calendar year and age, and
causing within-cohort changes of the  health indicator with
rate of change corresponding to the strength of this factors.

For illustrative purposes we have selected one-parameter case with
continuous, normally distributed parameter and with strength
numerically equal to rate of change.  While keeping model reasonably
realistic, these simplifications help to highlight  the key
properties of the dynamic model of population health and method of
its identification - the Dynamic Regression Method.

In the illustrative example,  we have shown that the Dynamic
Regression Method provides a sensible view on the BMI dynamics. It
reveals clear difference between the levels of the parameter and its
C-trends. From practical prospectives, it is C-trends, not levels,
which primarily seem to be modifiable by preventive activities or
involuntary changes affecting the population. It is worth noting
that outcomes from the DRM analysis serve as data for the next-level
analysis, involving other information and aiming at finding
reasonable explanation of the observed dynamics (diagnostic property
of DRM). One of the important complementary component for such an
analysis is dynamics of the population size ( we have developed a
modification of the DRM for that type of data, this is a subject for
one of the next publication). If there is significant migration "in"
or "out" of the study population, the observed effects could be
entirely or partially due to the population instability (health
selective effect). The outcomes from the DRM analysis could be used
straightforwardly for prediction of the age-specific profile of the
State Variable, say, for 5 year  period, by applying the C-trends at
the last year of the study period to the estimates of the
parameter's levels at that year. Such a projection will not cover
the cohorts, not included in the study age range at the last study
year.

Recall that this method has been developed  as an alternative to the
secular trends used so far. In this respect, it is worth noting that
the model presented here is characterized by local cohort trends
(C-trends), which have clear interpretation: changes in the State Variable of the same physical entity per time unit. If we will
formally calculate a characteristics resembling age-specific secular
trend, we will obtain  a difference between two different physical
entities (birth cohorts), caught occasionally at the moments of
measurement. Hence, it may behave quite arbitrarily. In other words,
in the view of the dynamic modeling approach,  secular trends do not
exist in nature. In one  special case only, when all the age
profiles of a State Variable are the same over calendar years
(stationary case), formally calculated secular trends will be equal
to zero at each age within the study age range. Only in that trivial
case, secular trends possess both, predictive and diagnostic power.However, even in this case, secular trends are kind of statistical fallacy, since  missing causality.

There are certain restrictions in using the current version of DRM
methods,  imposed by
the size of the problem, due to using matrix operations. Transfer to
the Bayesian Data Analysis and using Markov chain Monte Carlo
simulation methods \cite{Gel1} seems to be a solution for these
problems.

The  simplified dynamic equation used in the current model could  be
modified, accounting for the fact  that rate of change may  depend also on
the current level of the State Variable.

Finally, the most comprehensive model needs to be developed,
comprising multiple State Variables, and corresponding C-trends as a
linear functions of current State Variables. Such a model could be a
powerful practical tool for prediction of population health for about 5 year span.


\section*{Acknowledgements}
We thank The National FINRISK Study steering group for providing the data for illustrative
analysis.

\bibliography{vm15_110_bib}

\begin{thebibliography}{}

\bibitem[\protect\astroncite{Campbell}{1974}]{Campbell}
Campbell, S. (1974).
\newblock {\em Flaws and Fallacies in Statistical Thinking.}
\newblock Prentice Hall, Inc., Englewood Cliffs, NJ

\bibitem[\protect\astroncite{Chen et~al.}{2003}]{Chen2003}
Chen, X., Li, G., Unger, J.~B., Liu, X., and Johnson, C.~A. (2003).  Secular
  trends in adolescent never smoking from 1990 to 1999 in {C}alifornia: and
  age-period-cohort analysis.
\newblock {\em Am J Public Health}, 93:2099--104.

\bibitem[\protect\astroncite{Dobson et~al.}{1998a}]{dobsonst1}
Dobson, A.~J., Evans, A., Ferrario, M., Kuulasmaa, M., Moltchanov, V., Sans,
  S., Tunstall-Pedoe, H., Tuomilehto, J., Wedel, H., and {Yarnell J. for the
  WHO MONICA Project} (1998a).  Changes in estimated coronary risk in the
  1980s: data from 38 populations in the {WHO MONICA} project.
\newblock {\em Ann Med}, 30:199--205.

\bibitem[\protect\astroncite{Dobson et~al.}{1998b}]{dobsonst2}
Dobson, A.~J., Kuulasmaa, K., Moltchanov, V., Evans, A., Fortmann, S.~P.,
  Jamrozik, K., Sans, S., and { Tuomilehto J. for the WHO MONICA Project}
  (1998b).  Changes in cigarette smoking among adults in 35 populations in the
  mid-1980s.
\newblock {\em Tobacco Control}, 7:14--21.

\bibitem[\protect\astroncite{Ellner and Guckenheimer}{2006}]{Ellner}
Ellner, S.~P. and Guckenheimer, J. (2006).
\newblock {\em Dynamic Models in Biology}
\newblock Princeton University Press

\bibitem[\protect\astroncite{Gelman et~al.}{1995}]{Gel1}
Gelman, A., Carlin, J.~B., Stern, H.~S., and Rubin, D.~B. (1995).
\newblock {\em Bayesian Data Analysis}
\newblock Chapman \& Hall, London

\bibitem[\protect\astroncite{Gregg et~al.}{2005}]{Edward2005}
Gregg, E.~W., Cheng, Y.~J., Cadwell, B.~L., Flegal, K.~M., Narayan, K. M.~V.,
  and Williamson, D.~F. (2005).  Secular trends in cardiovascular disease risk
  factors according to body mass index in us adults.
\newblock {\em JAMA}, 293:1868 -- 74.

\bibitem[\protect\astroncite{Jaffe and Spirer}{1987}]{Jaffe}
Jaffe, A. and Spirer, H. (1987).
\newblock {\em Misused Statistics: Straight Talk for Twisted Numbers.}
\newblock Marcel Dekker, Inc., New York and Basel

\bibitem[\protect\astroncite{Kautiainen et~al.}{2002}]{Kautiainen2002}
Kautiainen, S., Rimpela, A.~H., Vikat, A., and Virtanen, S.~M. (2002).  Secular
  trends in overweight and obesity among finnish adolescents in 1977-1999.
\newblock {\em Int J Obes Relat Metab Disord}, 26:544 -- 52.

\bibitem[\protect\astroncite{Kuulasmaa et~al.}{2000}]{Kuulasmaa2000}
Kuulasmaa, K., Tunstall-Pedoe, H., Dobson, A., Fortmann, S., Sans, S., Tolonen,
  H., Evans, A., Ferrario, M., and for~the WHO MONICA~Project, J.~T. (2000).
  Estimation of contribution of changes in classic risk factors to trends in
  coronary-event rates across the {WHO MONICA Project} populations.
\newblock {\em Lancet}, 355:675--87.

\bibitem[\protect\astroncite{Luenberger}{1979}]{Luenberger}
Luenberger, D.~G. (1979).
\newblock {\em Introduction to Dynamic Systems: Theory, Models, and
  Applications}
\newblock John Wiley \& Sons, New York

\bibitem[\protect\astroncite{Mann}{2003}]{Mann2003}
Mann, C.~J. (2003).  Observational research methods. research design ii:
  cohort, cross sectional, and case-control studies.
\newblock {\em Emerg. Med.}, 20:54 -- 60.

\bibitem[\protect\astroncite{Moltchanov}{1993}]{Moltchanov1993}
Moltchanov, V. (1993).  The projection of the theory and methodology of the
  dynamic systems into epidemiological research.
\newblock {\em Can J Cardiol}, 9:88--89.

\bibitem[\protect\astroncite{Moltchanov et~al.}{1999}]{Moltchanov1999}
Moltchanov, V., Kuulasmaa, K., and {Torppa J. for the WHO MONICA Project}
  (1999).
\newblock {\em Quality assessment of demographic data in the {WHO MONICA
  Project}.}
\newblock Available as
  {http://www.ktl.fi/publications/monica/demoqa/demoqa.htm}

\bibitem[\protect\astroncite{Moltchanov and Mik'halskii}{2008}]{MolMic}
Moltchanov, V.~A. and Mik'halskii, A.~I. (2008).  Estimation of dynamics of
  risk factors by the dynamic regression method.
\newblock {\em Automation and Remote Control}, 69:125--140.

\bibitem[\protect\astroncite{Neumaier}{1999}]{Neumaier}
Neumaier, A. (1999).  Solving ill-conditioned and singular linear systems: A
  tutorial on regularization.
\newblock {\em SIAM Review}, 40:636--666.

\bibitem[\protect\astroncite{Newton and Motte.}{1995}]{Newton}
Newton, I. and Motte., A. (1995).
\newblock {\em The principia}
\newblock Great Minds New York : Prometheus Books
\newblock translation from Latin to English.

\bibitem[\protect\astroncite{Rao}{1973}]{Rao}
Rao, R.~C. (1973).
\newblock {\em Linear Statistical Inference and its Applications. Second
  edition}
\newblock John Wiley \& Sons, New York

\bibitem[\protect\astroncite{{SAS Institute Inc.}}{2004a}]{SAS9.1GRAPH}
{SAS Institute Inc.} (2004a).
\newblock {\em SAS/GRAPH\textsuperscript{\textregistered} 9.1 User's Guide}
\newblock Cary, NC: SAS Institute Inc.

\bibitem[\protect\astroncite{{SAS Institute Inc.}}{2004b}]{SAS9.1IML}
{SAS Institute Inc.} (2004b).
\newblock {\em SAS/IML\textsuperscript{\textregistered} 9.1 User's Guide}
\newblock Cary, NC: SAS Institute Inc.

\bibitem[\protect\astroncite{{SAS Institute Inc.}}{2004c}]{SAS9.1STAT}
{SAS Institute Inc.} (2004c).
\newblock {\em SAS/STAT\textsuperscript{\textregistered} 9.1 User's Guide}
\newblock Cary, NC: SAS Institute Inc.

\bibitem[\protect\astroncite{{SAS Institute Inc.}}{2011}]{SAS9.3L}
{SAS Institute Inc.} (2011).
\newblock {\em SAS\textsuperscript{\textregistered} Language Reference:
  Concepts 9.3}
\newblock Cary, NC: SAS Institute Inc.

\bibitem[\protect\astroncite{Strasak et~al.}{2007}]{Strasak}
Strasak, A.~M., Zaman, Q., Pfeiffer, K.~P., Göbel, G., and Ulmer, H. (2007).
  Statistical errors in medical research – a review of common pitfalls.
\newblock {\em SWISS MED WKLY}, 137:44--49.

\bibitem[\protect\astroncite{Tolonen et~al.}{2000}]{surveydb}
Tolonen, H., Kuulasmaa, K., and {for the WHO MONICA Project}, E.~R. (2000).
\newblock {\em MONICA population survey data book.}
\newblock Available as
  {http://www.ktl.fi/publications/monica/surveydb/title.htm}

\bibitem[\protect\astroncite{Tunstall-Pedoe et~al.}{2003}]{MoniMono}
Tunstall-Pedoe, H., editor. Prepared~by H~Tunstall-Pedoe, Kuulasmaa, K.,
  Tolonen, H., Davidson, M., and {Mendis S. with 64 other contributors for The
  WHO MONICA Project} (2003).
\newblock {\em MONICA Monograph and Multimedia Sourcebook.}
\newblock Geneva: World Health Organization

\end{thebibliography}
\bibliographystyle{astronm}

\newpage
\pagestyle{empty}

\end{document}